\documentstyle[prd,aps]{revtex}

\input epsf
\input psfig

\begin{document}

\thispagestyle{empty}

\draft

\title{New constraints on inflation from the cosmic microwave background}
\author{William H.\ Kinney\thanks{Electronic address: {\tt 
kinney@phys.ufl.edu}}}
\address{Department of Physics, University of Florida}
\address{P.O. Box 118440, Gainesville, FL 32611, USA}
\author{Alessandro\ Melchiorri\thanks{Electronic address: {\tt 
Alessandro.Melchiorri@roma2.infn.it}}}
\address{Dipartimento di Fisica, Universit\'a la Sapienza, Rome, Italy}
\author{Antonio\ Riotto\thanks{Electronic address: {\tt riotto@cibs.sns.it}}}
\address{Scuola Normale Superiore, Piazza dei Cavalieri 7}
\address{Pisa I-56126, Italy}

\date{\today}
\maketitle

\begin{abstract}
\baselineskip=15pt
The  recent data from the Boomerang and MAXIMA-1 balloon flights have 
marked the beginning of the precision era of Cosmic Microwave Background 
anisotropy (CMB) measurements. We investigate  the observational 
constraints from the current CMB anisotropy measurements on the  simplest 
inflation models, characterized by a single scalar field $\phi$, in the 
parameter space consisting of scalar spectral index $n_S$ and 
tensor/scalar ratio $r$. If we include constraints on the baryon density 
from big bang nucleosynthesis (BBN), we show that the favored inflationary 
models have negligible tensor amplitude and a ``red'' tilt, with a best 
fit of $n_S \simeq 0.93$, which is consistent with the simplest 
``small-field'' inflation models, but rules out large-field models 
at the $1\sigma$ level. Without including  BBN constraints, a  broader 
range of models are consistent with the data. The best fit (assuming 
negligible reionization) is a scale-invariant  spectrum, $n_S \simeq 1$, 
which includes large-field and hybrid scenarios. Large-field models 
(such as those typical of the chaotic inflation scenario) with tilt 
$n_S < 0.9$ are strongly disfavored in all cases.
\end{abstract}

\pacs{98.80.Cq,98.70.Vc,98.80.Es; SNS-PH-00-12}

\baselineskip=20pt

\section{Introduction}
\label{secintro}

It is commonly   believed that the Universe underwent an
early era of cosmological inflation. The flatness and the  
horizon
problems of the standard big bang
cosmology are elegantly solved if, during the evolution of
the early
Universe, the energy density is dominated by some  vacuum energy
and comoving scales grow
quasi-exponentially. The  prediction of the simplest models of
inflation
is a flat Universe, {\it i.e.} $\Omega_{tot}=1$ with great precision.

Inflation \cite{guth81} has also become the dominant 
paradigm for understanding the 
initial conditions for structure formation and for Cosmic
Microwave Background (CMB) anisotropy. In the
inflationary picture, primordial density and gravity-wave fluctuations are
created from quantum fluctuations ``redshifted'' out of the horizon during an
early period of superluminal expansion of the universe,
 where they
are ``frozen'' as perturbations in the background
metric\cite{muk81,hawking82,starobinsky82,guth82,bardeen83}. Metric 
perturbations at
the surface of last scattering are observable as temperature anisotropy in the
CMB.

The first and most 
impressive confirmation of the inflationary paradigm came when the CMB
 anisotropies were
 firmly detected by the COBE satellite in 1992 
\cite{smoot92,bennett96,gorski96}. Subsequently, it
became clear that  
the  
measurements of the spectrum of the CMB anisotropy can provide very detailed 
information
about the fundamental cosmological parameters \cite{cp}
and other crucial parameters  for particle physics \cite{pp}. This era of 
precision CMB anisotropy measurements has just begun with the results
of two balloon-borne experiments, Boomerang \cite{boom2000,boom2000a}  -- 
which was flown over Antarctica
in 1999 --  and MAXIMA-1 \cite{MAXIMAa,MAXIMAb}.  
The observed first acoustic peak at 
$\ell \sim 200$ (already detected by previous
experiments, see e.g. \cite{toco,boom97}) indicates that
the curvature of the Universe is consistent with flatness,
  thus providing another  convincing confirmation of the fact 
that the early Universe experienced
 a period of superluminal acceleration.

Despite the simplicity of the inflationary paradigm, 
the number of inflation models
that have been proposed in the literature is enormous 
\cite{lrreview}. 
Models have been invented which predict  non-Gaussian density
fluctuations\cite{nongaussianinflation}, isocurvature fluctuation
modes\cite{isocurvatureinflation}, and cosmic
strings\cite{stringinflation}. At present, inflation plus
topological defects is consistent with the Boomerang
data\cite{bouchet00} and  the evidence for non-Gaussianity in the COBE
maps\cite{evidencefornongaussianity} is tantalizing but not compelling. 
Furthermore, a small class of models predicting $\Omega_0<1$ (open
inflation models) have not survived the  current data.  

The goal of this paper is to   discuss the capability of the existing 
 CMB anisotropy  data  to discriminate among 
 the very simplest type of inflationary models: inflation involving a single
 scalar field (the inflaton)  
producing a flat Universe consistent with the
 observed first acoustic peak at $\ell \sim 200$.  The single-field models
are certainly the most popular 
among  models for cosmological inflation because  
they are    firmly rooted in modern
       particle theory and  have usually supersymmetry as a crucial
 ingredient \cite{lrreview}.

For single-field inflation models, the relevant parameter space 
for distinguishing among models is the plane defined by the scalar 
spectral index $n_S$ and the ratio of tensor to scalar fluctuations $r$.
 Recent work has shown that there is a kinematically favored region
 in the $r\,-\,n_S$ plane, independent of the details of the model
\cite{hoffman00}. Within that ``attractor'' region, different models make
 distinct predictions for $r$ and $n_S$. In this paper we show that
 the Boomerang and MAXIMA-1 data sets allow us to significantly 
constrain the allowed region in the $r\,-\,n_S$ plane and begin 
to rule out particular models of inflation. Such constraints are
 sensitive to our assumptions about other cosmological parameters,
 in particular the baryon density $\Omega_{\rm b}$ and the reionization
 optical depth $\tau_c$. If we allow both of these parameters to vary freely, a 
fairly large region in the $r\,-\,n_S$ plane is consistent with observation. A 
scale-invariant spectral index $n_S = 1$ is favored, and the models which 
predict both a tilted spectrum $n_S < 1$ and large tensor modes $r > 0.5$ are 
strongly disfavored. If, conversely, we choose to constrain $\Omega_{\rm b}$ 
using constraints from primordial nucleosynthesis, the allowed region in the 
parameter plane is tightened considerably. If we additionally assume negligible 
reionization optical depth, the data significantly favor a tilted spectrum $n_S 
< 1$ and negligible tensor modes.  

The paper is organized as follows. In Section II we describe the classes of 
inflation models we consider and their predictions for cosmological parameters 
which can be distinguished via CMB observations. In Section III we discuss the 
methods used for the CMB analysis. In Section IV we present the constraints from 
the combined Boomerang and MAXIMA-1 data sets. In Section V we discuss some 
general conclusions.

\section{The inflationary model space}
\label{seczoology}

In single-field models of inflation, 
the inflaton ``field'' need not be a fundamental field at all, of
course. Also, some ``single-field'' models require auxiliary fields. Hybrid
inflation models\cite{linde91,linde94,copeland94}, for example, require a second 
field to end inflation. What
is significant is that the inflationary epoch be described by a single
dynamical order parameter, the  inflaton field. 
 The metric perturbations created during inflation are of two types:
 scalar, or {\it curvature} perturbations, which couple to the 
stress-energy of matter in the universe and form the ``seeds'' 
for structure formation, and tensor, or gravitational wave 
perturbations.
 Both scalar and tensor perturbations contribute to CMB 
anisotropy. Scalar fluctuations can also be interpreted as 
fluctuations in the density of the matter in the universe. Most (but not all) 
inflationary models predict that these fluctuations are generated with 
approximately power law 
spectra\cite{muk81,hawking82,starobinsky82,guth82,bardeen83}
\begin{eqnarray}
P_{S}\left(k\right) \propto&& k^{n_{S} - 1},\cr
P_{T}\left(k\right) \propto&& k^{n_{T}}.
\end{eqnarray}
The spectral indices $n_{S}$ and $n_{T}$ are assumed to vary slowly or not at 
all with scale:
\begin{equation}
{d n_{S,T} \over d \log k} \simeq 0.
\end{equation}
In actuality, for a  spectral index sufficiently higher than unity, the 
power-law approximation becomes less and less accurate\cite{hoffman00,martin00}. 
Some particular inflation models predict running of the spectral index with 
scale
\cite{stewart97,stewart97a,copeland97,kinney97,covi98,kinney98,covi99,covi00} or 
even sharp features in the power spectrum\cite{featuresinpowerspectrum}. The 
observational consequences of scale dependence of the spectral index are 
considered in Ref. \cite{copeland97}. We do not consider such models here.  In 
principle, then, we have four independent observables with which to test 
inflation. One can be removed by considering the overall amplitude of CMB 
fluctuations: if the contribution
of tensor modes to the CMB anisotropy can be neglected, normalization to the
COBE four-year data gives\cite{bunn96,lyth96} $P_{S}^{1/2} = 4.8 \times
10^{-5}$. We then have three relevant observables: the tensor scalar ratio 
defined as a ratio of quadrupole moments,
\begin{equation}
r \equiv {C^{\rm Tensor}_2 \over C^{\rm Scalar}_2}
\end{equation}
and the two spectral indices $n_{S}$ and $n_{T}$. 

Even restricting ourselves to a simple single-field inflation scenario, the
number of models available to choose from is large \cite{lrreview}.
 It is convenient to define a general classification scheme, or ``zoology'' for 
models of inflation. We divide models into three general types: {\it 
large-field}, {\it small-field}, and {\it hybrid},  with a fourth 
classification, {\it linear} models, serving as a boundary between large- and 
small-field. A generic single-field potential  can be characterized by two 
independent mass scales: a ``height'' $\Lambda^4$, corresponding to the vacuum 
energy density during inflation, and a ``width'' $\mu$, corresponding to the 
change in the field value $\Delta \phi$ during inflation:
\begin{equation}
V\left(\phi\right) = \Lambda^4 f\left({\phi \over \mu}\right).
\end{equation}
Different models have different forms for the function $f$. The height $\Lambda$ 
is fixed by normalization, so the only free parameter is the width $\mu$. To 
create the observed flatness and homogeneity of the universe, we
require many e-folds of inflation, typically $N \simeq 50$. This figure varies
somewhat with the details of the model. During inflation, scales smaller than 
the horizon are ``redshifted'' to scales larger than the horizon. A comoving 
scale $k$ crosses the horizon during inflation $N\left(k\right)$ e-folds from 
the end of inflation, where $N\left(k\right)$ is given by\cite{lidsey97}
\begin{equation}
N(k) = 62 - \ln \frac{k}{a_0 H_0} - \ln \frac{10^{16}
        {\rm GeV}}{V_k^{1/4}}
        + \ln \frac{V_k^{1/4}} {V_e^{1/4}} - \frac{1}{3} \ln
        \frac{{V_e}^{1/4}}{\rho_{{\rm RH}}^{1/4}} \, .
\end{equation}
Here $V_k$ is the potential when the mode leaves the horizon, $V_e$ is the 
potential at the end of inflation, and $\rho_{{\rm RH}}$ is the energy density 
after reheating. Scales of order the current horizon size exited the horizon at 
$N\left(k\right) \sim 50 -70$. In keeping with the goal of discussing the most 
generic possible case, we will allow $N$ to vary within the range $50 \leq N 
\leq 70$ for any given model. Useful parameters for our purpose are the 
so-called {\it slow roll} parameters $\epsilon$ and $\eta$, which are related to 
the potential as\cite{lidsey97}:
\begin{equation}
\epsilon = {M_{Pl}^2 \over 16 \pi}
\left({V'\left(\phi\right) \over V\left(\phi\right)}\right)^2,
\end{equation}
and
\begin{equation}
\eta\left(\phi\right) = {M_{Pl}^2 \over 8 \pi}
\left[{V''\left(\phi\right) \over V\left(\phi\right)} - {1 \over 2}
\left({V'\left(\phi\right) \over V\left(\phi\right)}\right)^2\right].
\end{equation}
We can write the observable parameters $r$, $n_{S}$ and $n_{T}$ in terms of the 
model parameters $\epsilon$ and $\eta$\cite{turner93}:
\begin{eqnarray}
r =&& 13.6 \epsilon,\cr
n_{S} =&& 1 - 4 \epsilon + 2 \eta,\cr
n_{T} =&& - 2 \epsilon.
\end{eqnarray}
Note that $r$, $n_{S}$ and $n_{T}$ are not independent. The tensor spectral 
index and the tensor/scalar ratio are related as
\begin{equation}
n_{T} = - {1 \over 6.8} r,\label{eqconsistencyrelation}
\end{equation}
known as the {\it consistency relation} for inflation. (This relation holds
only for single-field inflation, and weakens to an inequality for inflation
involving multiple degrees of freedom\cite{polarski95,bellido95,sasaki96}.)
The relevant parameter space for distinguishing between inflation models is
then the $r\,-\,n_{S}$ plane.  Different classes of models are distinguished
by the value of the second derivative of the potential, or, equivalently, by
the relationship between the values of the slow-roll parameters $\epsilon$ and
$\eta$\footnote{The designations ``small-field'' and ``large-field'' can
  sometimes be misleading. For instance, both the  $R^2$ model
  \cite{starobinsky80} and the ``dual inflation'' model \cite{bellido98} are 
characterized by $\Delta \phi \sim m_{Pl}$, but are ``small-field'' in the sense 
that $\eta < 0 < \epsilon$, with $n_S < 1$ and negligible tensor modes.}. Each 
class of models has a different relationship between $r$ and $n_{S}$. For a more 
detailed discussion of these relations, the reader is referred to Refs. 
\cite{dodelson97,kinney98a}.  It should be noted that the accuracy of the slow 
roll approximation itself begins to break down for a spectral index 
significantly far from $n_S = 1$\cite{martin00}. First order in $\epsilon$ and 
$\eta$ is sufficiently accurate for the purpose of this analysis.

\subsection{Large-field models: $-\epsilon < \eta \leq \epsilon$}

Large-field models are potentials typical of the ``chaotic'' inflation 
scenario\cite{linde83}, in which the scalar field is displaced from the minimum 
of the potential by an amount usually of order the Planck mass. Such models are 
characterized by  $V''\left(\phi\right) > 0$, and $-\epsilon < \eta \leq 
\epsilon$. The generic large-field potentials we consider are polynomial 
potentials $V\left(\phi\right) = \Lambda^4
\left({\phi / \mu}\right)^p$,
and exponential potentials, $V\left(\phi\right) = \Lambda^4 \exp\left({\phi / 
\mu}\right)$. For the case of an exponential potential, $V\left(\phi\right) 
\propto \exp\left({\phi / \mu}\right)$, the tensor/scalar ratio $r$ is simply 
related to the spectral index as
\begin{equation}
r = 7 \left(1 - n_S\right).
\end{equation}  
This result is often incorrectly generalized to all slow-roll models, but is in 
fact characteristic {\it only} of power-law inflation. For inflation with a 
polynomial potential $V\left(\phi\right) \propto \phi^p$,  we again have $r 
\propto 1 - n_S$, 
\begin{equation}
r \simeq 7 \left({p \over p + 2}\right) \left(1 - n_S\right).
\end{equation}
so that tensor modes are large for significantly tilted spectra.

\subsection{Small-field models: $\eta < -\epsilon$}

Small-field models are the type of potentials that arise naturally
 from spontaneous symmetry breaking (such as the original models of ``new'' 
inflation \cite{linde82,albrecht82}) and from pseudo Nambu-Goldstone modes 
(natural inflation\cite{freese90}). The field starts from near an
 unstable equilibrium (taken to be at the origin) and rolls 
down the potential to a stable minimum. Small-field models are 
characterized by $V''\left(\phi\right) < 0$ and $\eta < -\epsilon$. Typically 
$\epsilon$ (and hence the tensor amplitude) is close to zero in small-field 
models. The generic small-field potentials we consider are of the form 
$V\left(\phi\right) = \Lambda^4 \left[1 - \left({\phi / \mu}\right)^p\right]$,
 which can be viewed as a lowest-order Taylor expansion of an arbitrary
 potential about the origin. The cases $p = 2$ and $p > 2$ have very different 
behavior. For $p = 2$,
\begin{equation}
r = 7 (1 - n_S) \exp\left[- 1 - N\left(1 - n_S\right)\right],
\end{equation}
where $N$ is the number of e-folds of inflation. For $p > 2$, the scalar 
spectral index is 
\begin{equation}
n_S \simeq 1 - {2 \over N} \left({p - 1 \over p - 2}\right),
\end{equation}
{\it independent} of $r$. Assuming $\mu < M_{\rm Pl}$ results in an upper bound 
on $r$ of
\begin{equation}
r < 7 {p \over N \left(p - 2\right)} \left({8 \pi \over N p \left(p - 
2\right)}\right)^{p / \left(p - 2\right)}.
\end{equation}

\subsection{Hybrid models: $0 < \epsilon < \eta$}

\begin{figure}
\psfig{figure=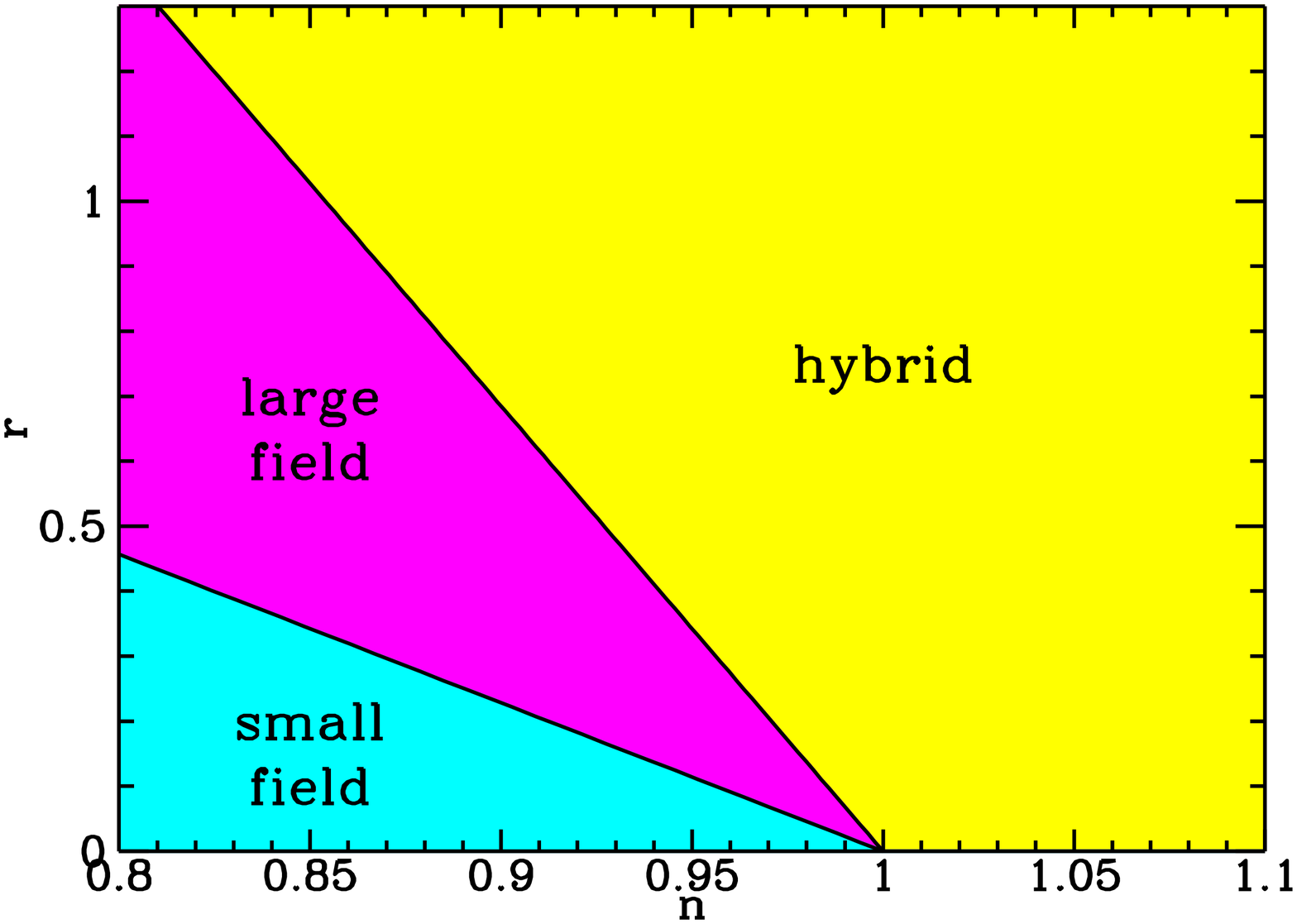,height=4.5in}
\caption{The parameter space divided into regions for small-field, large-field
and hybrid models. The linear case is the dividing line between large- and 
small-field.}
\label{figregions}
\end{figure}

The hybrid scenario\cite{linde91,linde94,copeland94} frequently appears in 
models which incorporate inflation into supersymmetry. In a typical hybrid 
inflation model, the scalar field responsible
for inflation evolves toward a minimum with nonzero vacuum energy. The end of 
inflation arises as a result of instability in a second field. Such models are 
characterized by $V''\left(\phi\right) > 0$ and $0 < \epsilon < \eta$. We 
consider generic potentials for hybrid inflation of the form $V\left(\phi\right) 
= \Lambda^4 \left[1 + \left({\phi / \mu}\right)^p\right].$ The field value at 
the end of inflation is determined by some other physics, so there is a second 
free parameter characterizing the models. Because of this extra freedom, hybrid 
models fill a broad region in the $r\,-\,n_S$ plane (Fig. \ref{figregions}). 
There is, however, no overlap in the $r\,-\,n_S$ plane between hybrid inflation 
and other models. The distinguishing feature of many hybrid models is a {\it 
blue} scalar spectral index, $n_S > 1$. This corresponds to the case $\eta > 2 
\epsilon$. Hybrid models can also in principle have a red spectrum, $n < 1$. 

\subsection{Linear models: $\eta = - \epsilon$}

Linear models, $V\left(\phi\right) \propto \phi$, live on the boundary between
large-field and small-field models, with $V''\left(\phi\right) = 0$ and $\eta = 
- \epsilon$. The spectral index and tensor/scalar ratio are related as:
\begin{equation}
r = {7 \over 3} \left(1 - n_S\right).
\end{equation}

This enumeration of models is certainly not exhaustive. There are a number of
single-field models that do not fit well into this scheme, for example
logarithmic potentials $V\left(\phi\right) \propto
\ln\left(\phi\right)$ typical of supersymmetry
\cite{lrreview}. Another example is potentials
with negative powers of the scalar field $V\left(\phi\right) \propto
\phi^{-p}$ used in intermediate inflation \cite{barrow93} and dynamical
supersymmetric inflation \cite{kinney97,kinney98}. Both of these cases require 
and auxilliary field to end inflation and are more properly categorized as 
hybrid models, but fall into the small-field region of the $r\,-\,n_S$ plane. 
However, the three classes
categorized by the relationship between the slow-roll parameters as $-\epsilon < 
\eta \leq \epsilon$ (large-field), $\eta \leq -\epsilon$ (small-field, linear), 
and $0 < \epsilon < \eta$ (hybrid), cover the entire $r\,-\,n_S$ plane and are 
in that sense complete.\footnote{Ref. \cite{kinney98a} incorrectly specified $0 
< \eta \leq \epsilon$ for large-field and $\eta < 0$ for small-field.} Figure 
\ref{figregions}\cite{dodelson97} shows the $r\,-\,n_S$ plane divided up into 
regions representing the large field, small-field and hybrid cases.

\section{CMB Analysis}
\label{secconstraints}

In this section we discuss the ability of current CMB data to 
place constraints on the inflationary parameter space. 
In particular, we wish to determine which regions of the inflationary 
model space are ruled out by the Boomerang and MAXIMA-1 data, and which 
regions are consistent with the observed data. 
We perform a likelihood analysis over multiple cosmological parameters 
and project likelihood contours onto the $r\,-\,n_S$ plane. 
The choice of parameters to be varied in the analysis is crucial,
 and different assumptions result in different constraints on the
 inflationary model space. In addition to $n_S$ and $r$, 
we also vary the Hubble constant $h$, the reionization
optical depth $\tau_c$, the amplitude of fluctuations, $C_{10}$,
(in units of $C_{10}^{COBE}$), the cold dark matter density 
$\Omega_{\rm M}$, the vacuum energy density $\Omega_\Lambda$, 
and the baryon density $\Omega_{\rm M}$, subject to the constraint 
of a flat universe $\Omega_{\rm total} = \Omega_{\rm B} + 
\Omega_{\rm CDM} + \Omega_\Lambda = 1$, consistent with 
the prediction of inflation. We also assume the consistency relation
 (\ref{eqconsistencyrelation}) between $n_T$ and $r$. We fix the number
of neutrino species at three and we take a gaussian prior
on the Hubble constant: $h= 0.65 \pm 0.2$.
Note that this approach is different from many CMB parameter
 estimation programs, which attempt 
to constrain cosmological parameters in a model independent fashion. 
Since our object here is to falsify models, we include as many model
 {\em dependent} constraints as possible and examine the 
consistency of the data with the predictions of the models.
 For example, this allows us to avoid questions of parameter 
mis-estimation arising from features in the primordial 
power spectrum\cite{kinney00}. 

The Boomerang and Maxima power spectra are estimated in
$12$ and $10$ bins respectively, spanning the range
$25 \le \ell \le 785$. In each bin, the spectrum is assigned
a flat shape, $\ell(\ell+1)C_{\ell}/2\pi=C_B$.
Following \cite{bjk00} we use the offset lognormal approximation
to the likelihood $L$. In particular we define \footnote{The
cross-correlation between bandpowers and the offset lognormal
corrections for the Boomerang and Maxima experiments 
are not yet public available. 
We tested the stability of our result including a $10 \%$ correlation
and with different lognormal distributions. We obtained near identical results
with the analysis of \cite{jaffe2k} where the comparison was
possible ($\Omega=1$, $r=0$).}:

\begin{equation}
-2{\rm ln} L
=(D_B^{th}-D_B^{ex})M_{BB'}(D_{B'}^{th}-D_{B'}^{ex}),
\end{equation}
\begin{equation}
D_{B}^{X}={\rm ln}(C_{B}^X+x_B),
\end{equation}
\begin{equation}
M_{BB'}=(C_B^{th}+x_B)F_{BB'}(C_{B'}^{ex}+x_{B'}),
\end{equation}
where $C_B^{th}$ ($C_B^{ex}$) is the theoretical (experimental)
band power, $x_B$ is the offset correction and $F_{BB'}$ is
the Gaussian curvature of the likelihood matrix at the peak.
Proceeding as in Refs. \cite{dk00,melk00}, we assign a likelihood to each
point in the parameter space and we find constraints
on all the parameters by finding the remaining ``nuisance''
parameters which maximize it.

Usually there are two maximization procedures. One
is based on a search algorithm through the second derivative
of the likelihood matrix\cite{dk00}. In this
approach the $C_{\ell}$'s are computed on the way,
without sampling the whole parameter space. The second approach
is based on building a database of $C_{\ell}$'s on a discretized
grid of the parameter space. The maxima values are then obtained
from the likelihood computed on the grid\cite{melk00,tegmark00}. We adopted
the database approach mainly for its robustness and the
possibility to use a 'true' marginalization method'.
The bulk of our results were evaluated on a database of 
models sampled as follows:

\begin{eqnarray}
&&\Omega_M=\Omega_{CDM}+\Omega_B=(0.10, 0.15,..., 1.1)\cr
&&\Omega_{\rm b}=(0.015,0.030,\ldots,0.20)\cr
&&\Omega_{\Lambda}=(0.0,\ldots,0.95)\cr
&&n_s=(0.50,0.52,\ldots,1.48,1.50)\cr
&&h=(0.25,0.30,\ldots,0.95)\cr
&&\tau_c=(0.0, 0.2, 0.4)\cr
&&n_t\ =-r / 6.8 \cr
&&r\ {\rm continuous}\cr
&&C_{10}\ {\rm continuous}.\nonumber
\end{eqnarray}

Two parameters in particular require careful consideration:
 the physical baryon density $\Omega_{\rm b} h^2$
 and the reionization optical depth $\tau_c$. 
Our constraints on the inflationary parameter space are 
strongly dependent on our choice of priors for $\Omega_{\rm b} h^2$.
 In particular, the suppressed second peak seen in the data requires
 either a baryon density too high to be consistent 
with big-bang nucleosynthesis (BBN) constraints\cite{tegmark00},
 or a significant ``red'' tilt, $n_S < 1$, if standard BBN is
assumed. 
 For completeness, we will consider separately $3$ cases: no prior
constraints on $\Omega_{\rm b} h^2$, 
$0.005 \le \Omega_{\rm b} h^2 \le 0.014$ and 
$0.016 \le \Omega_{\rm b} h^2 \le 0.021$ as suggested by the 
dichotomy in high and low deuterium measurements from distant 
quasars absorption line systems (see, e.g. Ref. \cite{burles99,espo00}).
 The second important parameter is the reionization optical depth $\tau_c$.
 It is known that there is a strong degeneracy between $\tau_c$
and the tensor/scalar ratio $r$, so that allowing 
for significant reionization strongly degrades the ability to distinguish 
tensor modes\cite{kinney98a}. Theoretical predictions, based on
the Press-Schecter formalism, show that reionization most
probably occurred after redshift $z \sim 50$\cite{tegsil},
with $z \sim 10-20$ for $\Lambda\ {\rm CDM}$ flat models\cite{balt98,haim98}. 
This roughly corresponds to an optical depth $0 \le \tau_c \le 0.2$\cite{grif99} 
.
Again, we consider separately 3 cases: no reionization,
 $\tau_c =0.2$ and $\tau_c =0.4$. 

\section{Results}

In the $r\,-\,n_S$-plane,  the parameter space is
divided into regions for small-field, large-field and hybrid
models as in Figure 1.
Figure 2 shows the $1\sigma$ ($\delta \chi^2 = 2.3$),
$2\sigma$ ($\delta \chi^2 = 6.0$), and $3\sigma$
($\delta \chi^2 = 9.2$) contours on the $r\,-\,n_S$
plane for the case of no reionization and no BBN prior.
 The various inflation models described in the last section and
 in Figure 1
are now plotted as labeled lines on the graph.
 (Note that the entire region to the right of the case of
the exponential potential is consistent with hybrid inflation models).
 The lines are obtained by varying a given parameter within a class
of models, {\it e.g.} for large-field models with a power-law potential
$\phi^p$ what varies is the power $p$.

The best model is nearly scale invariant, and the
tensor/scalar ratio $r$ is only weakly constrained.
Marginalizing over $r$ ($n_s$) we obtain also the $1-\sigma$
constraint : $0.94 < n_s < 1.07$ ($r < 0.65$).
 Hybrid and small-field models are consistent with the data,
but strongly tilted large-field models are in conflict with
 the CMB constraint to high significance.
 The $3\sigma$ contour on the $V\left(\phi\right) \propto \phi^p$
 models result in a constraint that $p < 5$ for $N = 50$ and
 $p < 8$ for $N = 70$. In the power-law inflation case,
 $V\left(\phi\right) \propto \exp\left(\phi / \mu\right)$,
the $3\sigma$ constraint corresponds to a lower limit
$\mu > 0.75 M_{\rm Pl}$.
The $1\sigma$ contour is marginally consistent with  $p = 2$ for $N = 50$ and 
requires $p < 3$ for $N = 70$ in the case of a polynomial potential,
and constrains $\mu \geq  M_{\rm Pl}$ in the power-law inflation case.

Figure 3 shows a similar plot when BBN constraints are included, still assuming 
negligible reionization. The contours in $r$ are significantly tighter than in 
the case with no BBN prior, and the best fit model is shifted toward tilted 
models: $0.88 < n_S < 0.98$ and $r \le 0.17$ for the low deuterium (LD) case ; 
$0.87 < n_S < 0.97$ and $r \le 0.16$ for the high deuterium (HD) case, with the 
scale invariant case just consistent with the data at $2\sigma$. (Fig. 3 shows 
the high deuterium case.) To $1\sigma$, large-field models are excluded 
entirely.  The only models within  $1\sigma$ are linear and small-field 
inflation. To $3\sigma$, the case of a polynomial potential is constrained to $p 
< 6$ for $N = 50$ and $p < 8$ for $N = 70$.

Figure 4 shows the constraints when a
 strong constraint on $\Omega_{\rm b}$ from BBN is
assumed (left to right)
and when the optical depth $\tau_c$ is increased.
Fixing the optical depth to $\tau_c \sim 0.2$ shifts the
likelihood towards ``blue'' tilted models ($n_S \ge 1$)
making the scalar invariant models consistent with the data even
when the BBN constraint are assumed
($1.02 < n_S < 1.28$ for no constraint, $0.95 < n_S < 1.05$
for both BBN constraints, all at $68 \%$ c.l.).
Increasing this parameter up to $\tau_c =0.4$ moves the $68 \%$
 likelihood contours in the hybrid models region
($1.15 < n_S < 1.30$, $1.06 < n_S < 1.14$ for LD and
$1.00 < n_S < 1.08$ for HD). These models are disfavored
by the dataset itself, being the best fit model at
$\Delta\chi^2 \sim 3$ from the corresponding best fit model
with $\tau_c=0$. Note in particular that large-field models with 
tilt $n_s < 0.9$ are strongly disfavored in all cases.
Removing the consistency relation and fixing $n_t=0$ does'nt
affect this conclusion and has small effect on the overall
result: for no BBN constraint and $\tau_c=0$ we
found $r < 0.67$ and $0.96 < n_s < 1.14$ (marginalized over
$r$).

\section{Conclusions}

With the release of data from the Boomerang and MAXIMA-1 balloon 
flights, it is possible for the first time to place significant constraints
 from the cosmic microwave background on the space of possible
 inflation models. The data in the region $\ell \ge 300$, in particular,
break the degeneracy between $n_S$ and $r$ on the first peak and produce
 a more stringent upper limit on $r$ for hybrid models respect to
 previous analysis\cite{melk99,zibi99,mikeh}.
In this paper we consider the observational constraints on the ``zoo'' of simple 
inflation models characterized by a single scalar field $\phi$ in the parameter 
space consisting of scalar spectral index $n_S$ and tensor/scalar ratio $r$. 
Different classes of models make distinct predictions for the relationship 
between $n_S$ and $r$ and can in principle be differentiated by use of CMB 
observations. While the current data is not good enough to rule out entire 
classes of models, it is sufficient to place significant constraints on model 
parameters. Combined with constraints on the baryon density from Big-Bang 
nucleosynthesis, we conclude that the favored inflationary models have 
negligible tensor amplitude and a ``red'' tilt, with a best fit of $r \le 0.17$ 
and $n_S = 0.93 \pm 0.05$ (assuming low deuterium and negligible reionization), 
consistent with the simplest ``small-field'' inflation models such as those 
arising from spontaneous symmetry breaking\cite{linde82,albrecht82} or from 
pseudo Nambu-Goldstone bosons (natural inflation\cite{freese90}). Models with 
strong reionization favor higher values of the spectral index consistent with 
hybrid inflation models. Without including BBN constraints, a broad range of 
models are consistent with the data, including large-field and hybrid scenarios, 
and the favored model is nearly scale-invariant, $n_S \simeq 1$. Future 
observations promise to significantly narrow the allowed regions of the 
parameter space, and will potentially make it possible to rule out entire 
classes of inflation models.

\section*{Acknowledgments}

We would like to thank Edward W. Kolb, Andrei Linde, David Lyth and
Jean Philippe Uzan for useful discussions. 
AM is grateful to Amedeo Balbi, Paolo de Bernardis, 
Ruth Durrer, Pedro Ferreira, Naoshi Sugiyama, Nicola Vittorio and 
the Boomerang Collaboration. WHK is supported in part by U.S. DOE grant 
DE-FG02-97ER-41029 at University of Florida.

\begin{figure}
\psfig{figure=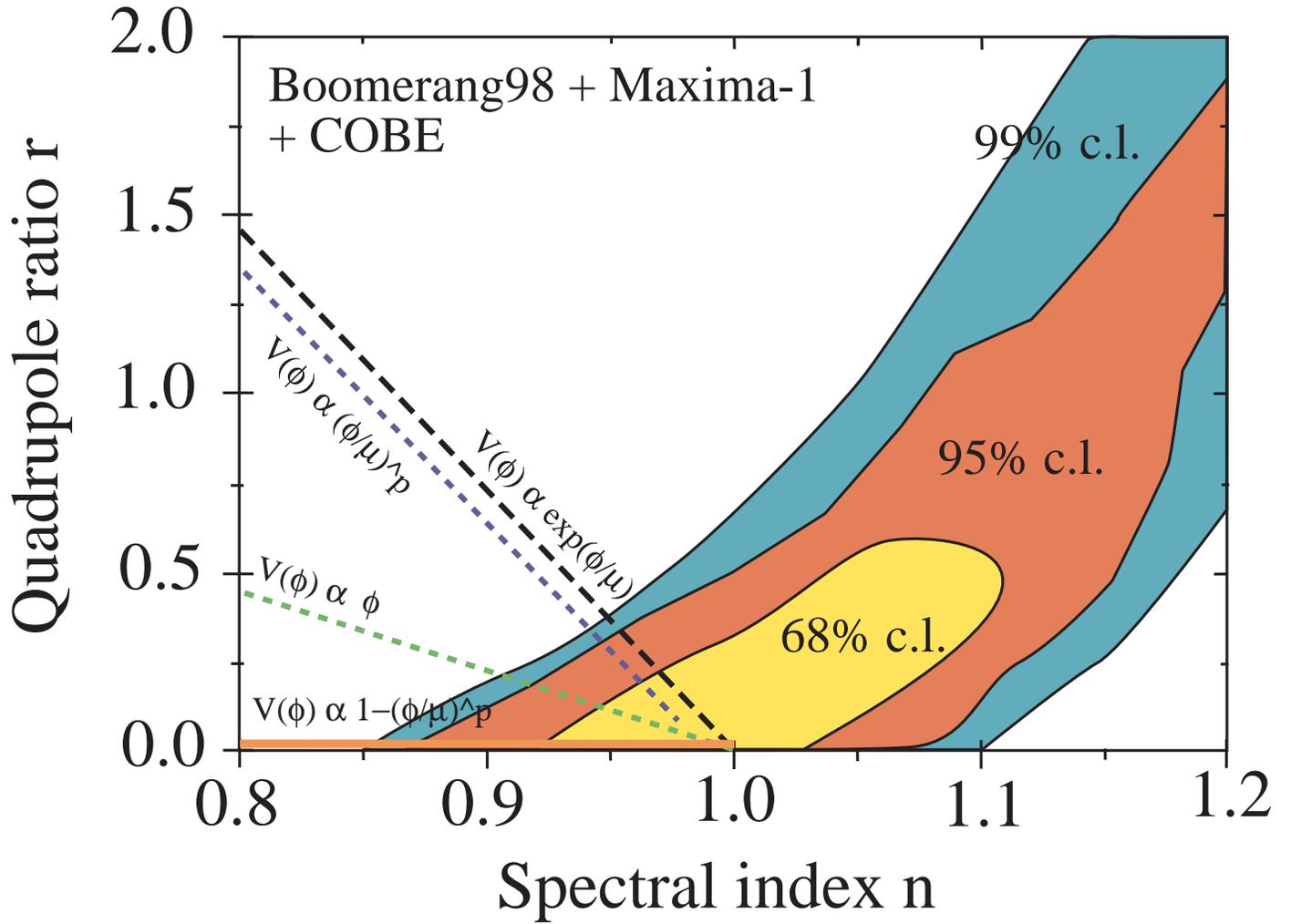,height=8.5in}
\caption{CMB constraints and inflation models for $\tau_c = 0$ and no BBN prior. 
The allowed contours are quite large but still exclude a significant portion of 
the inflationary model space.}
\end{figure}

\begin{figure}
\psfig{figure=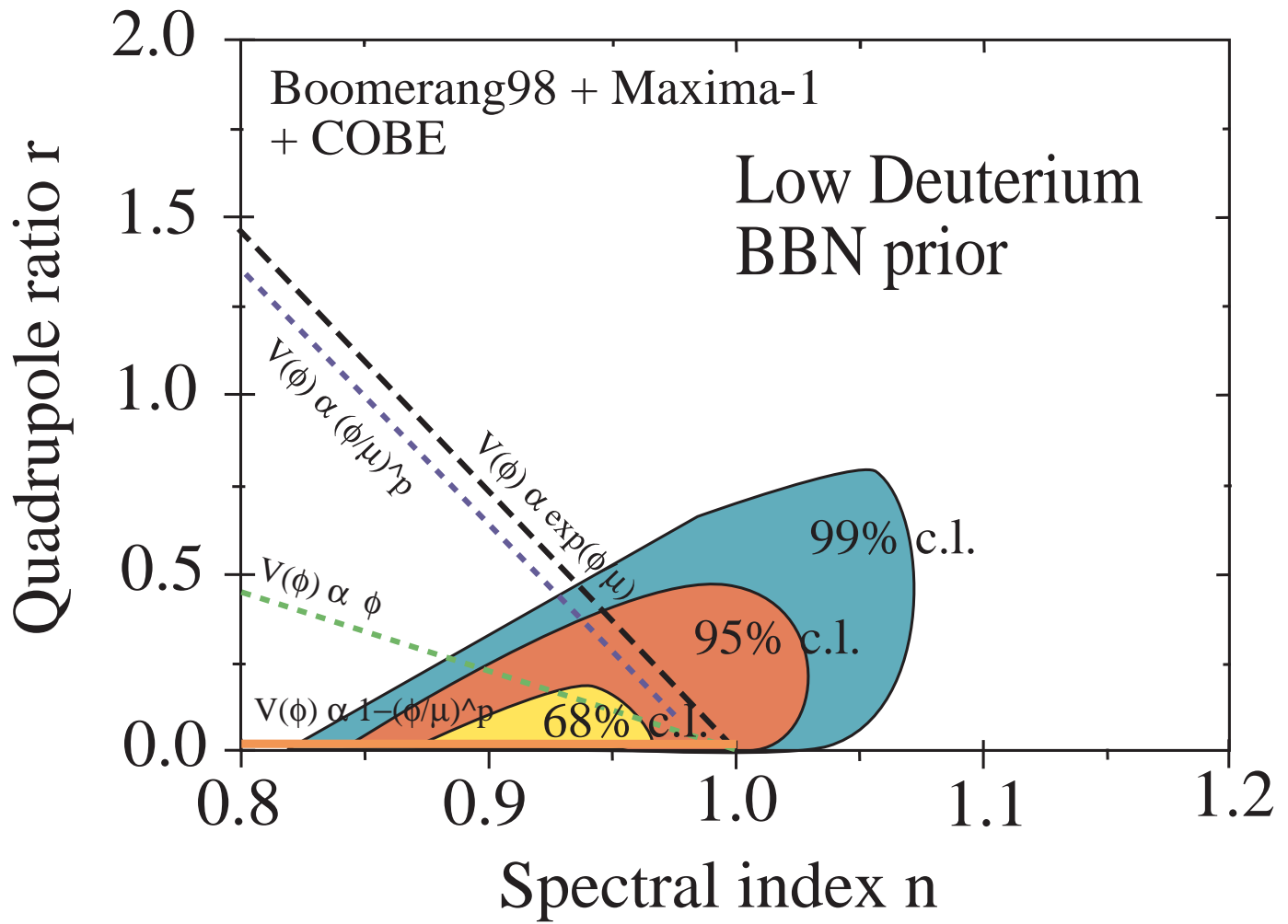,height=8.5in}
\caption{CMB constraints and inflation models for $\tau_c = 0$ and the low 
deuterium BBN prior, $0.016 \le \Omega_{\rm b} h^2 \le 0.021$. The contours are 
significantly tightened in the $r$ coordinate and now favor a tilted spectrum.}
\end{figure}

\begin{figure}
\psfig{figure=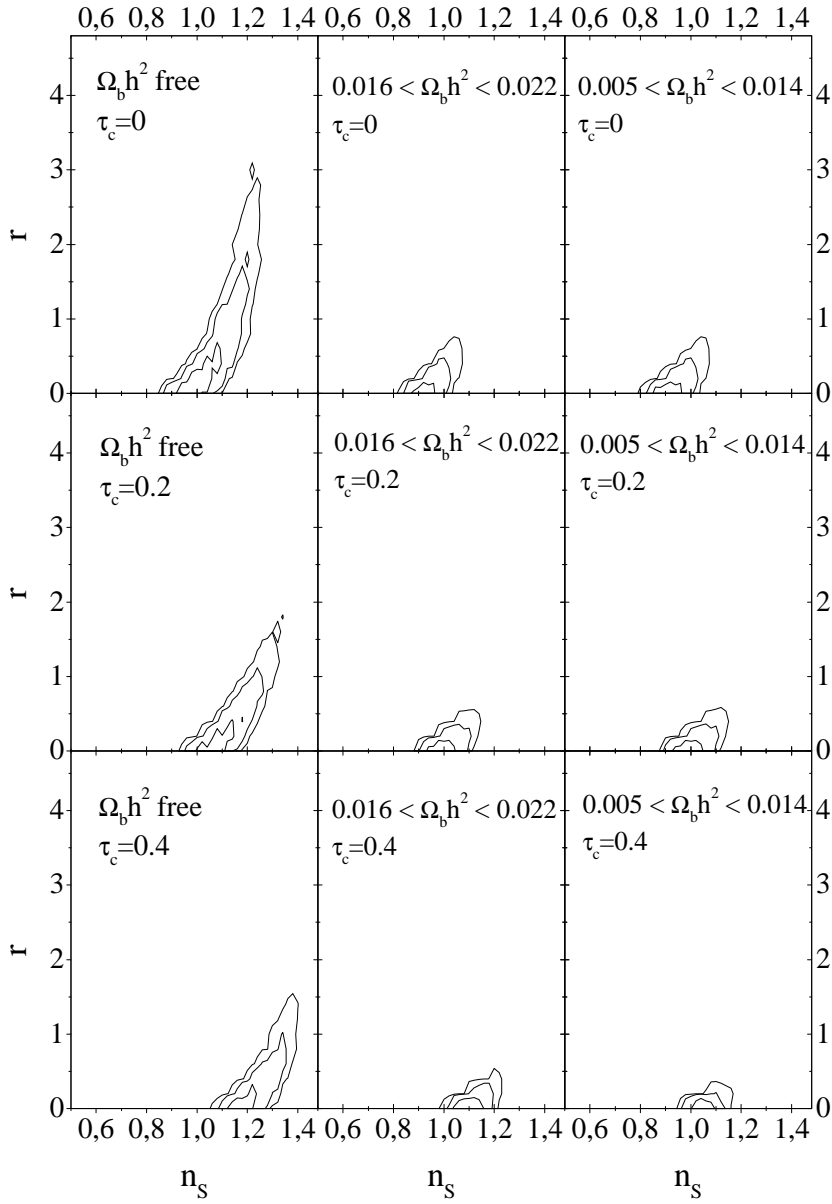,height=7.5in}
\caption{Contours for various values of $\tau_c$ and $\Omega_{\rm b} h^2$. In 
general, BBN constraints tighten the contours in the $r$ direction, and 
reionization shifts the favored range of $n_S$ to the right, favoring hybrid 
models for large $\tau_c$. Strongly tilted large-field models are excluded in 
all cases.}
\end{figure}

\end{document}